\documentclass[11pt,a4paper,english,amssymb,nofootinbib,superscriptaddress,showpacs]{revtex4}
\usepackage{lmodern}

\usepackage[T1]{fontenc}
\usepackage[latin9]{inputenc}
\usepackage{slashed}
\usepackage{babel}
\usepackage{amsmath}
\usepackage{slashed}
\usepackage{amssymb}
\usepackage{graphicx}
\usepackage{epstopdf}
\usepackage{subfigure}
\usepackage{esint}
\usepackage[unicode=true,pdfusetitle,
bookmarks=true,bookmarksnumbered=false,bookmarksopen=false,
breaklinks=false,pdfborder={0 0 1},backref=false,colorlinks=false]
{hyperref}

\makeatletter
\@ifundefined{textcolor}{}
{%
	\definecolor{BLACK}{gray}{0}
	\definecolor{WHITE}{gray}{1}
	\definecolor{RED}{rgb}{1,0,0}
	\definecolor{GREEN}{rgb}{0,1,0}
	\definecolor{BLUE}{rgb}{0,0,1}
	\definecolor{CYAN}{cmyk}{1,0,0,0}
	\definecolor{MAGENTA}{cmyk}{0,1,0,0}
	\definecolor{YELLOW}{cmyk}{0,0,1,0}
}

\usepackage{latexsym}\usepackage{bm}

\makeatother

\makeatother

\begin{document}
	\begin{flushright}{
			MIT-CTP-4834\\}
		
	\end{flushright}
\title{A Noncompact Weyl-Einstein-Yang-Mills Model: A Semiclassical Quantum Gravity}         

\author{Suat Dengiz}

\email{sdengiz@mit.edu, suatdengiz@yahoo.com}  

\affiliation{Center for Theoretical Physics,\\ 
	Massachusetts Institute of Technology, Cambridge MA 02139 USA}  

\date{\today}

\begin{abstract}                                         

We construct and study perturbative unitarity (i.e., ghost and tachyon analysis) of a $3+1$-dimensional \emph{noncompact} Weyl-Einstein-Yang-Mills model. The model describes a local noncompact Weyl's scale plus $SU(N)$ phase invariant Higgs-like field, conformally coupled to a generic Weyl-invariant dynamical background. Here, the Higgs-like sector generates the Weyl's conformal invariance of system. The action does not admit any dimensionful parameter and genuine presence of de Sitter vacuum spontaneously breaks the noncompact gauge symmetry in an analogous manner to the Standard Model Higgs mechanism. As to flat spacetime, the dimensionful parameter is generated within the dimensional transmutation in quantum field theories, and thus the symmetry is radiatively broken through the one-loop Effective Coleman-Weinberg potential. We show that the mere expectation of reducing to Einstein's gravity in the broken phases forbids anti-de Sitter space to be its stable vacua. The model is unitary in de Sitter and flat vacua around which a massless graviton, $N^2-1$ massless scalar bosons, $N$ massless Dirac fermions, $N^2-1$ Proca-type massive Abelian and non-Abelian vector bosons are generically propagated. 
\end{abstract}
\maketitle 
\section{Introduction}                                  
The recent discovery of gravitational wave has shown us one more time why Einstein's gravity deeply deserves to be the only universally acknowledged gravity theory. As is well-known, the theory was constructed on a novel geometrical pattern possessing a non-linear relation with the matter sector through Einstein field equation. Here, in the geometry sector, the metric is the \emph{only} dynamical variable that governs type of geometry and affine dynamics of manifolds. This unique property of metric in Einstein's gravity is the corollary of imposed constraints, that is the torsionlessness and metric compatibility, on a generic connection. With those constraints, the degrees of freedom (DOF) coming from torsion and non-metricity are automatically ruled out, and thus the only solutions that comprise the Levi-Civita connection are picked up as viable solutions of the theory. As for the generic case, this obviously does not have to be the situation and one has to allow for all the other DOF to get a well-behaved larger geometrical representation of gravitational force. This will inherently upgrade the fundamental behavior of Einstein's gravity \cite{Eddington, Schrodinger1, Schrodinger2, Hehl}. The necessity of readdressing these disregarded DOF particularly arises due to the UV problem of theory: recall that Einstein's gravity possesses Newton's constant which has mass dimension $-2$ as coupling constant. Having the dimensionful coupling constant particularly causes troubles in the perturbative study of theory. More precisely, as one moves ahead of the one loop self interactions of \emph{pure} theory in the radiative aspect, due to being a dimensionful parameter, Newton's constant necessitates infinite number of counter-terms for the renormalization. Thus, since the catastrophic infinities cannot be regulated with an acceptable number of counter-terms, the theory inevitably turns out to be non-renormalizable \cite{tHooftVeltman}. At this stage, it seems to be natural to think that the theory may not provide a consistent quantum gravity in this perspective. Or instead, to get a perturbatively compatible larger gravity theory, one might go back and remove the restrictions that Einstein initially imposed, and thus allow all (or some) of the above-mentioned forsaken DOF. In this regard, the partial stretching of keeping torsionlessness intact but allowing non-metricity is particularly attractive because it results in a habitat for local scale-invariant field theories. This sectional relaxation in the connection is called the Weyl's approach in which the rigid scale-invariance that requires the conformal flatness due to the Lorentz-invariance is promoted to a \emph{local} scale-invariance in order to get the Poincare-invariant field theories in \emph{arbitrary} curved spacetimes \cite{Weyl1918, Weyl1919, Weylbook, oraif, Iorio}. As is common knowledge, the local scale symmetry does not permit any dimensionful parameters in theories. Therefore, by assuming it to be a genuine symmetry of early universe, then the dimensionful parameters, such as Newton's constant, Higgs mass etc., can emerge only if this symmetry is lost. Thus, since special relativity also indicates that the masses of particles become meaningless at the extremely high energies, it is logical to expect from a complete quantum gravity theory to possess at least this symmetry. Recently, there appear several papers on the topic: see for example \cite{Jimenez1} for an interesting work on an $n$-dimensional extension of Gauss-Bonnet gravity theory in Weyl's geometry that intriguingly leads to the vector-tensor Horndeski interaction in $n=3+1$-dimensions. See also \cite{Tanhayindim} for the Weyl-gauging of $n$-dimensional quadratic curvature gravity theories in which it is shown that the Weyl-invariant Einstein-Gauss-Bonnet theory is the only unitary combination, and \cite{JackiwNewSet, Jimenez2, Yuan, Oliva, Maki} for some other related works. For the integration of local scale symmetry to Standard Model, see \cite{tHooft}. On the other hand, even though there are few works concerning to the quantizations of Weyl-gauged gravity models, there seems there is a manifest gap in this research area. See for example \cite{Shapiro, Percacci} for one-loop beta function calculations for particular Weyl-invariant gravity theories and \cite{Abhinav} for a recent work on quantization of a Weyl-invariant gravity model coupled to a Stuckelberg photon via background field method.

In the light of above discussion, here we will conjecture that it is the Higgs-like field that converts the Levi-Civita connection into non-metricit connection and accordingly construct a $3+1$-dimensional \emph{noncompact} Weyl-Einstein-Yang-Mills model in this paper. Here, the Higgs-like sector will generate the local Weyl's conformal invariance of system as expected. To supply the Weyl's symmetry from Higgs-like sector, we will choose the Higgs-like field to initially be an element of $SU(N)$ in the adjoint representation as in the Georgi-Glashow model \cite{Glashw}\footnote{Recall that the Georgi-Glashow model admits the $SU(5)$ gauge group and is the simplest Grand Unified Theory of fundamental particles which leads to the unique cosmic inflation \cite{Guth, Linde, AlbSte, GuthPi}.}, which will later be used to construct the Weyl's scalar and gauge fields. Subsequently, we will show that the noncompact Weyl plus $SU(N)$ symmetry will be spontaneously broken in de Sitter vacuum in an analogous manner to the Higgs mechanism \cite{Higgs, Engler, Anderson, Goldstone, Nambu1, Nambu2} and radiatively broken via the one-loop Effective Coleman-Weinberg potential in flat vacuum\cite{ColemanWeinberg}\footnote{See \cite{DengizTekin} for a similar work for particular Weyl-gauged higher curvature gravity theories.}. We will demonstrate that the model fails to be unitary in anti-de Sitter vacuum. In addition to this, we will also study the tree-level perturbative unitarity of model in de Sitter background. We will show that the model is unitary in de Sitter and flat vacua about which a massless graviton, $N^2-1$ massless scalar bosons, $N$ massless Dirac fermions, $N^2-1$ Proca-type massive Abelian and non-Abelian vector bosons are generically propagated.

The lay-out of paper is as follows: In Sec.II, we give the fundamental properties of Weyl's gauging by focusing on its implementation to common samples. In Sec.III, we study a particular embedding of Weyl's symmetry into $SU(N)$ and construct the $3+1$-dimensional noncompact Weyl-Einstein-Yang-Mills model. Here, we also evaluate the corresponding field equations and study the emerging symmetry breaking mechanisms either for de Sitter and flat vacua. In Sec.IV, we perform the tree-level unitarity of model around de Sitter vacuum and find the relevant unitary regions for parameters. In Sec.V, we summarize our results. In the Appendices, we respectively give the field equations of noncompact model, the basics of adjoint Higgs-like field of $SU(N)$ and the second order expansion of fundamental curvature tensors.  

\section{The Local Weyl's Gauging}              
Historically, the concept of non-metricity in connection goes back to Hermann Weyl's 1918 attempt of reconciliation of gravity with electromagnetism \cite{Weyl1918, Weyl1919, Weylbook}. Although the attempt did not provide a legitimate unification, his idea of gauging systems by virtue of compensating vector potentials has remained intact\footnote{In fact, due to Einstein's critic, Weyl's idea had been put aside until London's work in 1927 in which he showed that quantum theory also acquires a similar symmetry \cite{London}.}. The most promising property of Weyl's method is that it elevates the rigid scale-invariance to a \emph{local} scale-invariance in order to get Poincare-invariant field theories in \emph{arbitrary} curved spacetimes\footnote{The geometrical interpretation of Weyl's scaling is as follows: recall that in Einstein's gravity, as one parallel transports a given vector around a closed curve in (pseudo) Riemannian geometry, the direction of a vector changes due to curvature. As for Weyl's gravity, this parallel transportation changes not only the direction but also the magnitude of vector.}. To recapture basics of Weyl's symmetry, we will take a look at its application in the foremost $n$-dimensional field theories in this section. Later, we will fix the background to be $3+1$-dimension. (In doing so, we will mainly follow \cite{DengizTekin}. For a comprehensive understanding of Weyl's gauging, see also \cite{oraif, Iorio, Jimenez1}). Therefore, let us first notice that in order for the scalar field theory    
\begin{equation}
S_{\Phi}=- \frac{1}{2}\int d^n x \sqrt{-g} \, g^{\mu \nu} \partial_\mu \Phi \, \partial_\nu \Phi,
\label{weylscalkin}
\end{equation}
to be invariant under the following local Weyl's transformations in a generic dynamical curved background
\begin{equation}
g_{\mu\nu} \rightarrow g^{'}_{\mu\nu}=e^{2 \sigma(x)} g_{\mu\nu}, \hskip 1 cm \Phi \rightarrow \Phi^{'} =e^{-\frac{(n-2)}{2}\sigma(x)} \Phi,
\label{weytrnsf}
\end{equation}
where $\sigma(x)$ is any point-dependent function, one needs to change the ordinary partial derivative with the gauge-covariant derivative defined as  
\begin{equation}
\tilde{D}_\mu \Phi =\partial_\mu \Phi -\frac{n-2}{2} E_\mu \Phi  , \hskip 1 cm  \tilde{D}_\mu g_{\alpha \beta} =\partial_\mu g_{\alpha\beta} + 2 E_\mu g_{\alpha \beta}.
\label{weylgagcovder}
\end{equation}
Here $E_\mu$ is the corresponding Weyl's gauge field and transforms as follows
\begin{equation}
E_\mu \rightarrow E^{'}_\mu = E_\mu - \partial_\mu \sigma(x).
\label{weylgagfldtransf}
\end{equation}
Consequently, with all these set-ups, one arrives at 
\begin{equation}
 \tilde{D}_\mu g_{\alpha \beta} \rightarrow ( \tilde{D}_\mu g_{\alpha \beta})^{'}=e^{2 \sigma(x)} \tilde{D}_\mu g_{\alpha \beta} , \hskip 1 cm \tilde{D}_\mu \Phi \rightarrow  (\tilde{D}_\mu \Phi)^{'}=e^{-\frac{(n-2)}{2}\sigma(x)} \tilde{D}_\mu \Phi ,
\end{equation}
which manifestly provide the local gauge-invariance. As to Maxwell-type field theory, the action fails to be Weyl-invariant even if the field strength tensor $F_{\mu\nu}=\partial_\mu E_\nu-\partial_\nu E_\mu$ respects the symmetry\footnote{Note that this is also valid for the non-Abelian field-strength tensor.}. Accordingly, using a properly tuned compensating scalar field, one gets the Weyl-invariant extension of Maxwell-like field theory as follows
\begin{equation}
S_{E_\mu} =  - \frac{1}{2} \int d^n x \sqrt{-g}\,\, \Phi^{\frac{2(n-4)}{n-2}} F_{\mu \nu} F^{\mu \nu}.
\label{maxwell} 
\end{equation}
Note that the changes coming from the local transformations of scalar and tensor parts cancel out each others and thus Eq.(\ref{maxwell}) preserves its structure. Finally, one needs to upgrade the Levi-Civita connection to a proper larger non-metricit one in order to integrate the Weyl's symmetry to gravity. Referring \cite{DengizTekin} for details, let us note that the desired connection actually reads 
\begin{equation}
	\tilde{\Gamma}^\lambda_{\mu\nu}=\frac{1}{2}g^{\lambda\sigma} \Big ( \tilde{D}_\mu g_{\sigma\nu}+\tilde{D}_\nu g_{\mu\sigma}
	-\tilde{D}_\sigma g_{\mu\nu} \Big ),
\end{equation}
with which one gets the Weyl-invariant Riemann tensor
\begin{equation}
\begin{aligned}
\tilde{R}^\mu{_{\nu\rho\sigma}} [g,E]&=\partial_\rho \tilde{\Gamma}^\mu_{\nu\sigma}-\partial_\sigma \tilde{\Gamma}^\mu_{\nu\rho}
+ \tilde{\Gamma}^\mu_{\lambda\rho} \tilde{\Gamma}^\lambda_{\nu\sigma}-\tilde{\Gamma}^\mu_{\lambda\sigma} \tilde{\Gamma}^\lambda_{\nu\rho} \\
& =R^\mu{_{\nu\rho\sigma}}+\delta^\mu{_\nu}F_{\rho\sigma}+2 \delta^\mu{_[\sigma} \nabla_{\rho]} E_\nu 
+2 g_{\nu[\rho}\nabla_{\sigma]} E^\mu \\
& \hskip 3.2 cm+2 E_[\sigma   \delta_{\rho]}\,^\mu E_\nu + 2 g_{\nu[\sigma}  E_{\rho]} E^\mu  +2 g_{\nu[\rho} \delta_{\sigma]}\,^\mu  E^2 , 
\label{weinvriem}
\end{aligned} 
\end{equation}
where $ 2 E_{[ \rho} E_{\sigma]} \equiv E_\rho E_\sigma -  E_\sigma E_\rho$ and $E^2= E_\mu E^\mu$. Observe that by assuming the ordinary gauge covariant derivative of gauge field to be  $\tilde{D}_{\mu} E_\nu \equiv \partial_\mu E_\nu - E_\mu E_\nu$ and then letting its relation with spacetime covariant derivative to be ${\tilde{\mathcal{D}}}_{\mu} E_\nu \equiv \nabla_\mu E_\nu - E_\mu E_\nu$, one could write Eq.(\ref{weinvriem}) in a more compact form. In fact, we will benefit from this structure in the construction of noncompact gauge covariant derivative of gauge field in Sec.IV. Note also that due to the non-metricity in connection, the Weyl-gauged Riemann tensor does not possess the symmetries of ordinary Riemann tensor. Subsequently, the contraction of Eq.(\ref{weinvriem}) yields the Weyl-invariant Ricci tensor as follows 
\begin{equation}
\begin{aligned} 
\tilde{R}_{\nu\sigma} [g,E]&= \tilde{R}^\mu{_{\nu\mu\sigma}}[g,E] \\
&=R_{\nu\sigma}+F_{\nu\sigma}-(n-2)\Big [\nabla_\sigma E_\nu - E_\nu E_\sigma +E^2  g_{\nu\sigma} \Big ]-  g_{\nu\sigma}\nabla \cdot E,
\end{aligned}
\end{equation}
where $\nabla \cdot E \equiv \nabla_\mu  E^\mu$. Finally, the Weyl-gauged Ricci scalar becomes
\begin{equation}
\tilde{R}[g,E]=R-2(n-1)\nabla \cdot E-(n-1)(n-2) E^2,
\label{weyltransricciscalar}
\end{equation}
which transforms as follows
\begin{equation}
\tilde{R}[g, E] \rightarrow (\tilde{R}[g,E])^{'} = e^{-2 \sigma (x) }  \tilde{R}[g, E].
\end{equation}  
Thus, unlike Riemann and Ricci tensors, Ricci scalar is \emph{not} invariant under the Weyl's transformations. Therefore, here one also needs to consider a properly tuned compensating scalar field to get Weyl-invariant extension of Einstein's gravity. It is actually straightforward to show that the desired extension reads
\begin{equation}
\begin{aligned}
S&= \int d^n x \sqrt{-g} \, \Phi^2 \, \tilde{R}[g, E] \\
&= \int d^n x \sqrt{-g} \, \Phi^2 \, \Big [R-2(n-1)\nabla \cdot E-(n-1)(n-2) E^2 \Big].
\label{eh}
\end{aligned}
\end{equation}
As a cursory look, let us notice the following point: as is mentioned hitherto, the Weyl's gauging approach is an alternative and more generic way of providing conformal symmetry to appropriate field theories. Hence, it is natural to expect from the resident theories in Weyl's geometry to give the similar results obtained via the sister conformal theories in Einstein's geometry. In this aspect, the recent very interesting result of the vanishing of Noether current for Weyl's symmetry in conformally coupled scalar-tensor theory (CCSTT) shown by Jackiw and Pi \cite{JackiwPi} should also have a legitimate explanation in the Weyl's geometry context. To find a potentially affirmative bond between two frames for this specific case, let us now dwell on the condition(s) with which one will get the CCSTT in the Weyl's geometry, too. It is actually manifest that the bare Weyl-invariant Einstein-Hilbert action without any kinetic term for the gauge field fulfills this job. That is, note that by varying Eq.(\ref{eh}) with respect to the Weyl's gauge field, one gets  
\begin{equation}
E_{\mu }= \frac{2}{n-2}\partial_{\mu} \ln \Phi,
\label{puregageh}
\end{equation} 
whose re-insertion into Eq.(\ref{eh}) yields the CCSTT \cite{DengizTekin}  
\begin{equation}
S=\int d^n x \sqrt{-g}\Big [ \Phi^2 R+4\frac{(n-1)}{n-2} \partial_\mu \Phi \, \partial^\mu \Phi  \Big].
\label{scaltenswey}
\end{equation}   
Note that this elimination at the action level is allowed due to the constraint equation coming from non-propagating vector field as in \cite{Faddeev}\footnote{Notice that although the scalar field in Eq.(\ref{eh}) also seems to be a non-propagating DOF at first sight, a similar elimination cannot be done for the scalar field because its dynamic becomes manifest during CCSTT in Eq.(\ref{scaltenswey}).}. With this, as is also studied in \cite{ccstt1, ccstt2, ccstt3, ccstt4} in which the above reduction is entitled "Weyl integrable space-time", the scalar field in the ordinary CCSTT in fact acquires a \emph{geometrical} interpretation. Note also that this result is not valid if one assigns a proper Maxwell-type kinetic term to the gauge field. Thus, as a side comment, the above vanishing of the vector field leading to CCSTT in Weyl's geometry might be the core reason of why the Noether current for the Weyl's symmetry of CCSTT in the Einstein's geometry vanishes \cite{JackiwPi}. Of course, this is solely a cursory analysis and requires a comprehensive study in order to make a decisive conclusion\footnote{See also \cite{Oda1, Oda2} for similar works in which it is shown that the Noether current for the Weyl's symmetry in Weyl transverse (WTDiff) gravity (in unimodular gravity) vanishes, too.}. (See \cite{Campigotto} for a related work in which the conserved quantities associated to conformal and diffeomorphism symmetries for gauged version of conformal gravity are discussed in details.) Observe that, unlike the above situation, the Weyl's gauge field that is used in ensuing section is highly non-linear and thus dynamical. Finally, note that as the scalar field is freezed, Eq.(\ref{scaltenswey}) reduces to pure Einstein theory. In this respect, one also needs to allow for the Weyl-invariant scalar potential (that is, $ \Phi^{\frac{2n}{n-2}} $) to capture cosmological Einstein's theory as the scalar field is freezed properly. In this case, the full Weyl-invariant scalar action will become
\begin{equation}
S_\Phi=- \frac{1}{2}\int d^n x \sqrt{-g}\Big (\tilde{D}_\mu \Phi \tilde{D}^\mu\Phi +\nu \, \Phi^{\frac{2n}{n-2}}\Big ),
\label{scalarwithpot}
\end{equation}
where $\nu \ge 0$ is a dimensionless coupling constant ensuring that the potential admits a ground state.

\section{Extension of $SU(N)$ via Weyl's Symmetry and $3+1$-Dimensional Noncompact Weyl-Einstein-Yang-Mills Model} 
As was mentioned in the introduction, it is aimed to construct such a viable Weyl-invariant gravity plus Higgs-like model in which the existing Higgs-like sector will inherently generate the local conformal symmetry of whole system in this work. This naturally necessitates the Higgs-like field to also possess the Weyl's gauge symmetry. However, recall that the Higgs-like field has only $SU(N)$ as gauge symmetry and one needs to find a legitimate integration of Weyl's symmetry to $SU(N)$ so that the Higgs-like field will ultimately acquire both of the gauge symmetries at the same time. That is, the arising larger symmetry group will provide the Higgs-like field to describe the phase and scale symmetries simultaneously. In such a reconciliation, the most critical point will admittedly be the representation that Higgs-like field belongs. That is, the Higgs-like field has to be selected in such a way that the emergent larger group will not have any problem in suppling the Weyl's symmetry to the noncompact model. In this respect, since the Higgs-like field of $SU(N)$ in the fundamental representation would end up with inconsistency in the Weyl-invariant extension of gravity sector, let us choose the one that pertains to the \emph{adjoint representation}\footnote{See Appendix B for the fundamentals of adjoint Higgs-like field of $SU(N)$.} and subsequently impose its magnitudes in the generator bases to behave as Weyl's scalar field as follows   
\begin{equation}
\varphi^a \rightarrow \varphi^{a'}= e^{-\sigma(x)} \varphi^a,
\label{weylcompsca}
\end{equation}
which automatically induces the following modification in the group transformation
\begin{equation}
U \rightarrow {\cal U}=U \,\, e^{-\sigma(x)}.
\label{defgroptranfsg}
\end{equation}
Note that by expanding Eq.(\ref{defgroptranfsg}) in the generator bases, one can easily show that the emerging transformation parameter contains a real and a complex part. Here, the group theoretical interpretation of the above-made reconciliations of symmetries is that with the imposed condition in Eq.(\ref{weylcompsca}), the compact $SU(N)$ has actually been extended to the noncompact $SL(N, \mathbb{C})$ which has $2N^2-2$ generators. Thus, denoting the generators of $SL(N, \mathbb{C})$ as $\{ K^a \cdots \}$, they can be expressed via complexification of $SU(N)$ \cite{Fuchs, Barut, Hsu} as follows\footnote{In describing fundamentals of $SL(N, \mathbb{C})$, we use identical notations followed in \cite{Hsu, Dengizsun}. Notice that save for the similarity in notation, the model we are studying in this paper is completely different from the ones studied in these papers.}  
\begin{equation}
\{ K^a \cdots \} \equiv \{ T^a \cdots, \mbox{i} T^a \cdots \}, \qquad a=1,2,...,N^2-1.
\label{genarncompcompl}
\end{equation}   
In accordance with this, the corresponding noncompact gauge field $A_\mu$ can be written as a complexification of a non-Abelian gauge field ($B_\mu$) and a gauge covariant field ($C_\mu$)\footnote{Here, by gauge covariant field, we mean the vector field which transforms as $C_\mu \rightarrow C^{'}_\mu={\cal U} C_\mu {\cal U}^{-1}$ \cite{Nair}.}:
\begin{equation}
 A_\mu \equiv B_\mu+\mbox{i} C_\mu.
 \label{decnongfld}
\end{equation}
Note that the Higgs-like field transforms in agreement with the adjoint representation of $SL(N, \mathbb{C})$ as
\begin{equation}
\varphi \rightarrow \varphi^{'}={\cal U} \varphi {\cal U}^{-1}.
\label{noncmpscfldtrnsf}
\end{equation}
As to the local noncompact gauge invariance of scalar field theory, one needs to assume the following noncompact gauge covariant derivative defined in the adjoint representation
\begin{equation}
{\cal D}_\mu \varphi \equiv \partial_\mu \varphi- \mbox{i} g [ A_\mu, \varphi].
 \label{noncompgagcovder}
 \end{equation}
Here, the gauge field transforms as follows
\begin{equation}
A^\mu \rightarrow A^{'}_\mu= {\cal U} A_\mu {\cal U}^{-1}+\frac{1}{\mbox{i} g} (\partial_\mu {\cal U}){\cal U}^{-1},
\label{nongagtrnfsm}
\end{equation}
with which one finally gets
\begin{equation}
  {\cal D}_\mu \varphi \rightarrow {\cal D}^{'}_\mu \varphi = {\cal U} {\cal D}_\mu \varphi {\cal U}^{-1}.
\end{equation}
Notice that due to Eq.(\ref{genarncompcompl}), all the above settings can be expanded in the generator bases of $SU(N) $. This is particularly essential in the construction of kinetic term for the noncompact gauge field. To see this, let us first note that $ A_\mu $ can be recast as follows 
\begin{equation}
 A_\mu \equiv A^a_\mu T^a =(B^a_\mu+ \mbox{i} C^a_\mu)T^a,
 \label{decompnncmpcgf}
\end{equation}
where, unlike the ordinary compact case, the magnitudes are complex. Later, one needs to assume the following field-strength tensor
\begin{equation}
 F_{\mu\nu}=\partial_\mu A_\nu-\partial_\nu A_\mu- \mbox{i}g[A_\mu, A_\nu] \hskip 1cm \mbox{with} \hskip 1cm F_{\mu\nu} \rightarrow F^{'}_{\mu\nu}={\cal U} F_{\mu\nu} {\cal U}^{-1},
 \label{kinnoncmp}
\end{equation}
and then search for a convenient kinetic term for $ A_\mu $. Observe that Eq.(\ref{kinnoncmp}) can be written as $ F_{\mu\nu}=F_{\mu\nu}^a T^a$ where the magnitudes will read
\begin{equation}
 F^a_{\mu\nu}=\partial_\mu A^a_\nu-\partial_\nu A^a_\mu + gf^{abc}A^b_\mu A^c_\nu,
 \label{expnonfstts} 
\end{equation}
which, with $A^a_\mu=B^a_\mu + \mbox{i} C^a_\mu$, can be recast as follows
\begin{equation}
 F^a_{\mu\nu}={\cal B}^a_{\mu\nu}+\mbox{i} {\cal C}^a_{\mu\nu},
 \label{fstnoncompcast2} 
\end{equation}
where
\begin{equation}
 \begin{aligned}
{\cal B}^a_{\mu\nu}&= \partial_\mu B^a_\nu-\partial_\nu B^a_\mu+g f^{abc} (B^b_\mu B^c_\nu-C^b_\mu C^c_\nu) \\
 {\cal C}^a_{\mu\nu}&= \partial_\mu C^a_\nu-\partial_\nu C^a_\mu+g f^{abc} (B^b_\mu C^c_\nu+C^b_\mu B^c_\nu).
 \label{fstnoncompcast3} 
 \end{aligned}
\end{equation}
To conclude the discussion, let us first note that a naive attempt of getting canonically normalized Yang-Mills-type kinetic term 
\begin{equation}
 {\cal L}_{A_\mu}=-\frac{1}{4} \mbox{Tr}(F_{\mu\nu}F^{\mu\nu}),
 \label{kinymus}
\end{equation}
would apparently lead to the violation of unitarity. Therefore, one needs to look for another combination of field strength tensor in order to resolve this problem. In this respect, let us notice that although the following kinetic term 
\begin{equation}
	\begin{aligned}
	 {\cal L}_{A_\mu} \sim \mbox{Tr}(F_{\mu\nu}F^{+\mu\nu}) \sim {\cal B}^a_{\mu\nu}{\cal B}^{a\mu\nu}+{\cal C}^a_{\mu\nu}{\cal C}^{a\mu\nu} ,
 \label{kintermforgag}
	\end{aligned}
\end{equation}
seems to provide a unitary model, it fails to be invariant under the Weyl-Yang-Mills (WYM) transformation. Since the kinetic term in Eq.(\ref{kintermforgag}) is compatible with the unitarity but breaks down during the WYM transformation, one can actually focus on the problematic part and try to resolve the obstacles arising throughout the transformation, and thus get a proper WYM-invariant extension of Eq.(\ref{kintermforgag}). Whether or not there is any other alternative way, to our knowledge in the literature as in, for example, \cite{Hsu}, this can at least be achieved with the help of fermionic field as follows: let us consider the Dirac field $\psi$ which transforms according to the fundamental representation according to 
\begin{equation}
\psi \rightarrow \psi^{'}= {\cal U} \psi, \hskip 1 cm  ( {\cal D}_\mu \psi ) \rightarrow ( {\cal D}_\mu \psi )^{'} = {\cal U} ({\cal D}_\mu \psi ),
\end{equation}
where ${\cal D} =\partial_\mu \psi-\mbox{i} g A_\mu \psi$. Here, note that the Dirac Lagrangian 
\begin{equation}
{\cal L}_{Dirac}=\bar{\psi} \mbox{i} \gamma^\mu {\cal D}_\mu \psi,
\end{equation}
is not WYM-invariant. To resolve this problem, one can recast the gamma matrices as 
\begin{equation}
\gamma_\mu(x) \rightarrow \Gamma_\mu (x) \qquad \mbox{such that} \qquad \Gamma_\mu(x) \rightarrow \Gamma^{'}_\mu(x) =({\cal U}^{+})^{-1}\Gamma_\mu(x) {\cal U}^{-1}, 
\end{equation}
which generically requires an additional field. In this respect, we assume this extra field to be a function of the components of Higgs-like field and thus define
\begin{equation}
\Gamma_\mu(x)=\gamma^\mu \Theta \,[\varphi^a(x)],
\end{equation}
with which the WYM-invariant Dirac theory becomes
\begin{equation}
S_{Dirac} = \eta \int d^4 x \sqrt{-g} \, (\varphi^a)^2 \bar{\psi} i \Gamma^\mu {\cal D}_\mu \psi.
\end{equation}
In this case, the WYM-invariant kinetic term will read \cite{Hsu}
\begin{equation}
\begin{aligned}
\mbox{Tr} (F^{+}_{\mu\nu} \Theta F^{\mu\nu} \Theta^{-1} )&=-2 F^{+a}_{\mu\nu} F^{a\mu\nu}+\frac{4 \Theta^a \Theta^b}{{\bf \Theta}^2} F^{+a}_{\mu\nu} F^{b\mu\nu} \\
&= -2 \Big ({\cal B}^a_{\mu\nu}{\cal B}^{a\mu\nu}+{\cal C}^a_{\mu\nu}{\cal C}^{a\mu\nu}\Big)+\frac{4 \Theta^a \Theta^b}{{\bf \Theta}^2} \Big ({\cal B}^a_{\mu\nu}{\cal B}^{b\mu\nu}+{\cal C}^a_{\mu\nu}{\cal C}^{b\mu\nu}\Big).
\label{kintermforgagttttt}
\end{aligned}
\end{equation}
Note that without assuming the existence of Dirac theory, one would not achieve to build this WYM-invariant kinetic term. See Sec.IV for the perturbative unitarity of $3+1$-dimensional non-compact Weyl-Einstein-Yang-Mills model in which it is shown that Eq.(\ref{kintermforgagttttt}) accurately provides a tree-level unitary model.

Now that we have extended the local gauge symmetry of Higgs-like field to $SL(N, \mathbb{C})$ (i.e., local Weyl plus $SU(N)$ symmetry), we can skip to formulate the noncompact model. In doing so, the Higgs-like sector is required to ultimately generate the Weyl-invariance of system. But, note that the above set-ups cannot do the job unless a proper Weyl's gauge field is also defined. Actually, there is only one possible candidate in the above-stated construction to be interpreted as the Weyl's gauge field. It is the gauge covariant field $C^a_\mu$. However, since its pure form would cause problem in providing the local conformal invariance to gravity sector of the noncompact model, one needs to look for a viable combination of $C^a_\mu$ as Weyl's gauge field. Here, the crucial point is to find the most suitable combination. Although we give the detailed explanation below, here let us quote the final result: one can easily show that with the following particular \emph{superposition} of all real components $\varphi^a$ and $C^a_\mu$ as Weyl's gauge field  
\begin{equation}
	E_\mu=g f^{abc} C^a_\mu \varphi^b (\varphi^c)^{-1},   
	\label{criticalrelbtwrl}
\end{equation}
the Higgs-like sector ultimately provides the local conformal symmetry to the entire system coherently and simultaneously. Here, $(\varphi^a)^{-1}$ is simply the inverse of magnitude of Higgs-like field in $a^{th}$ generator basis. Observe that all the generator indices are completely contracted and since the terms $g, f^{abc}, \varphi^a $ and $C^a_\mu$ are real variables, the constructed Weyl's gauge field is a \emph{real} field. As is given below, this choice actually arises due to the above-given demands on the model. Note also that in order for the specific choice of Weyl's gauge field to satisfy the Weyl's gauge transformation in Eq.(\ref{weylgagfldtransf}), one needs to select $\sigma(x)$ according to
\begin{equation}
\sigma(x)=- g f^{abc} f^{klm} \int dx^{\mu} C^a_\mu (x) \varphi^b (x) (\varphi^c)^{-1}(x) \, w^k(x)w^l(x)T^m+H,  
\end{equation}	
where $H$ is an arbitrary constant. Consequently, by keeping in mind the basics of Weyl-invariance in Sec.II and accordingly by using the magnitudes of adjoint Higgs-like field in the generator bases (that is, $\varphi^a$'s) and specific choice of Eq.(\ref{criticalrelbtwrl}) as Weyl's scalar and gauge fields respectively, one can easily show that the most general action for $3+1$-dimensional \emph{noncompact} Weyl-Einstein-Yang-Mills model becomes  
\begin{equation}
	\begin{aligned}
		{\cal S}_{nWEYM} = \int d^4 x \sqrt{-g} \, \bigg \{& \alpha (\varphi^a)^2  \tilde{R}[g_{\mu\nu}, E_\mu] 
	+\beta ({\cal D}_\mu \varphi^a)({\cal D}^\mu \varphi^a)^{+}+ \gamma (\varphi^a)^4+\eta (\varphi^a)^2 \bar{\psi}_b \, \mbox{i} (\slashed{\cal D} \psi )^b \\
&+\sigma \Big[-2 \Big ({\cal B}^a_{\mu\nu}{\cal B}^{a\mu\nu}+{\cal C}^a_{\mu\nu}{\cal C}^{a\mu\nu}\Big)+\frac{4 \Theta^a \Theta^b}{{\bf \Theta}^2} \Big ({\cal B}^a_{\mu\nu}{\cal B}^{b\mu\nu}+{\cal C}^a_{\mu\nu}{\cal C}^{b\mu\nu}\Big) \Big]\bigg \},
		\label{maineqn}
	\end{aligned}
\end{equation}
where $\alpha, \beta, \gamma, \eta$ and $\sigma$ are arbitrary dimensionless couplings. In this case, the $3+1$-dimensional Weyl-extended Ricci scalar reads\footnote{With Eq.(\ref{criticalrelbtwrl}), the $3+1$-dimensional gauge covariant derivative of metric and Weyl's scalar fields in Eq.(\ref{weylgagcovder}) respectively become
\begin{equation}
\tilde{D}_\mu g_{\alpha\beta}=\partial_\mu g_{\alpha\beta}+2g f^{abc} C^a_\mu \varphi^b (\varphi^c)^{-1} g_{\alpha\beta}, \hskip 0.7 cm 
\tilde{D}_\mu \Phi=\partial_\mu \Phi-g f^{abc} C^a_\mu \varphi^b (\varphi^c)^{-1} \Phi, 
\end{equation}
where all the representation indices are closed.}  
\begin{equation}
\begin{aligned}
	\tilde{R}[g_{\mu\nu}, E_\mu] &=R-6 g f^{abc} \nabla_\mu C^{a\mu} \varphi^b (\varphi^c)^{-1} \\
	&\hskip 0.82 cm -6 g^2 f^{abc} f^{klm} C^a_\mu \varphi^b (\varphi^c)^{-1} \, C^{k\mu} \varphi^l (\varphi^m)^{-1},
	\label{manfeq}
\end{aligned}
\end{equation}
and the noncompact gauge covariant derivative is  
\begin{equation}
{\cal D}_\mu \varphi^a =\partial_\mu \varphi^a +g f^{abc} A^b_\mu \varphi^c=\partial_\mu \varphi^a +g f^{abc} B^b_\mu \varphi^c+\mbox{i} g f^{abc} C^b_\mu \varphi^c.
\end{equation}
Notice that as is required by the local conformal-invariance, the model does not contain any dimensionful parameter. Here, due to the stability analysis in next section, we avoid using the canonically normalized coefficients at this level. Regarding to Eq.(\ref{maineqn}), one should note that in addition to the noncompact kinetic term for scalar field, one could also add a separate kinetic term for the Weyl's gauge field defined in Eq.(\ref{weylgagcovder}). But this would not be an economical way. Rather, bearing in mind that we want the local scale-invariance to be supplied by the fundamental tools of adjoint Higgs-like sector, then one can easily show that with the specific choice of Weyl's gauge field in Eq.(\ref{criticalrelbtwrl}) and the following additional relation 
\begin{equation}
f^{abc} \Big [ (\partial_\mu \varphi^a) B^{b\mu} \varphi^c+ \frac{1}{2} C^a_\mu \varphi^b (\varphi^c)^{-1} \partial^\mu (\varphi^d)^2 \Big]=0,
\label{nonababgaufield}
\end{equation}
the noncompact kinetic term of scalar field in Eq.(\ref{maineqn}) describes all the dynamics of both Weyl's scale and $SU(N)$ phase symmetries coherently and simultaneously.

Let us now note that as the scalar field is freezed to its VEV as $\varphi^a=\left< \varphi^a_{vac} \right>$, the noncompact Weyl-Einstein-Yang-Mills model reduces to the ordinary Einstein's gravity with an effective Newton's constant $ \kappa = \left< \varphi^a_{vac} \right>^{-1} $. (Observe that in this case, Newton's constant gets contribution from all the VEV of scalar fields in all the generator bases.) At this stage, one naturally comes up with the question of whether this limit emerges as the vacuum solution of noncompact model or not, that is, whether the noncompact Weyl plus $SU(N)$ symmetry is broken by vacua or not. To this end, one shall compute the field equations. Referring Appendix A for the derivations of field equations, let us go ahead and analyze them in constant curvature vacua. To do so, let us first notice that fixing the noncompact field-strength tensor as $F^a_{\mu\nu}=0$ and particularly choosing $C^a_\mu=B^a_\mu=0$ will prevent the violation of local Lorentz-invariance of vacua. Thereupon, freezing the scalar field as $\varphi^a=\left< \varphi^a_{vac} \right>$ and using   
\begin{equation}
 R_{\mu\nu\alpha\beta}= \frac{\Lambda}{3} (g_{\mu\alpha} g_{\nu\beta}-g_{\mu\beta} g_{\nu\alpha}), \hskip 0.7 cm R_{\mu\nu}=\Lambda g_{\mu\nu}, \hskip 0.7 cm  R=4 \Lambda,
 \label{curvconstspc}
 \end{equation}
 the gauge field equations in Eq.(\ref{feqnonabefld})-Eq.(\ref{feqabefld}) vanish, whereas the tensor and scalar field equations in Eq.(\ref{feqofmetrcs})-Eq.(\ref{feqscfld}) yield  
\begin{equation}
 \Lambda=-\frac{\gamma \left< \varphi^a_{vac} \right>^2}{2 \alpha}, 
 \label{vfeqdd}
\end{equation}
which is the relevant \emph{vacuum field equation}. Note that here one has two cases: one can suppose the VEV of scalar field is known and then calculate the cosmological constant or vice versa. Firstly, notice that for the first option, only de Sitter spacetime ($\Lambda <0$) is allowed in order to get Einstein gravity in the vacua ($\alpha >0$) that admits a ground state ($\gamma >0$). Secondly, for the second option, Eq.(\ref{vfeqdd}) turns into
\begin{equation}
\left< \varphi^a_{vac} \right>=\pm \sqrt{-\frac{2 \alpha}{\gamma} \Lambda},
\end{equation}
where, due to the positivity of VEV of scalar field, the negative solution is automatically ruled out for both de Sitter ($\Lambda <0$) and anti-de Sitter ($\Lambda > 0$) spaces. Observe also that as in the first option, here the positivity and reality of Newton's constant which are required to obtain Einstein gravity in vacua allows only de Sitter spacetime. Thus, anti-de Sitter background is inevitably omitted in both cases.

As for flat vacua ($\Lambda=0$), the vacuum field equation in Eq.(\ref{vfeqdd}) yields $\left< \varphi^a_{vac} \right>=0$ and thus the noncompact symmetry remains unbroken. In order to resolve this problem, one could add an extra term that would lead to the spontaneous breaking of symmetry in vacua. Alternatively, one can analyze the model radiatively and check if there comes any dimensionful parameter at loop-level: as is well-known, the usual Coleman and Weinberg calculation \cite{ColemanWeinberg} is performed for massless $ \Phi^4$ interaction in $3+1$-dimensional flat space which, save for the generic group structure, is exactly our situation. Even though (or at least to our knowledge) there isn't a complete Coleman-Weinberg computation for generic $SU(N)$, let us notice that in accordance with the spontaneous symmetry breaking in the Georgi-Glashow model \cite{GGSBCM1, GGSBCM2, GGSBCM3, GGSBCM4, GGSBCM5, GGSBCM6}\footnote{That is, $SU(5) \rightarrow SU(3)_C \times SU(2)_I \times U(1)_Q \rightarrow SU(3)_C \times U(1)_Q$ where $SU(3)_C$, $SU(2)_I$ and $U(1)_Q$ stand for the color, isospin and hyper-charge gauge groups which respectively correspond to Strong, Weak and Electromagnetic interactions.} or Standard Model Electroweak sector via the Coleman-Weinberg mechanism \cite{ColemanWeinberg}, here the one-loop Effective scalar potential with generic structure
\begin{equation}
{\cal V}_{eff}=\alpha_1+ \alpha_2 \, (\varphi^a_{vac})^4 \Big[\log \frac{\varphi^a_{vac}}{ \left< \varphi^a_{vac} \right>}+\alpha_3\Big ], 
\end{equation}
also shifts the minima and generates a non-zero dimensionful parameter (that is, VEV of scalar field) and thus breaks the noncompact gauge symmetry. Note that due to our current aim of determining the viable symmetry-breaking mechanism in flat background, the explicit values of $\alpha_1, \alpha_2$ and $\alpha_3$ are not necessary. Apparently, one also needs to evaluate the contributions coming from graviton and gauge fields to scalar loops in order to have the full one-loop Effective potential. But, this is beyond the scope of paper and will presumably change only the numerical values as in for example \cite{Abbott}.   
  
\section{Tree-level Stability and Particle Spectrum}                          
As was seen in Sec.III, the natural expectation of getting Einstein's theory in the broken phase of noncompact Weyl-Einstein-Yang-Mills model allows only de Sitter spacetime to be its viable constant curvature vacuum. However, this cannot be taken as a conclusive result unless a detailed stability analysis is also performed. To clarify this point at least semi-classically, here we will study the tree-level perturbative unitarity of model by determining if there are ghost and tachyon-free parameter regions for the particles propagated around de Sitter vacuum. In what follows, we will utilize the background field method (BFM) \cite{tHooftVeltman} which suggests to expand actions up to quadratic order in fluctuations to obtain the lowest order quantum contributions to classical background solutions. Therefore, in order to apply the BFM to our model, let us now suppose that in de Sitter vacua, we have 
\begin{equation}
\begin{aligned}
  g_{\mu\nu} =\bar{g}_{\mu\nu}, \qquad \varphi^a_{vac} \equiv \left< \varphi^a_{vac} \right>,  \qquad \psi^a_{vac}=0, \qquad B^{a\mu}_{vac} = 0 , \qquad C^{a\mu}_{vac} = 0.
\label{vacstab} 
\end{aligned}  
\end{equation}
Recall that gauge fields are set to zero in order to prevent vacua to choose certain directions and hence break the local Lorentz-invariance. Here, $\left< \varphi^a_{vac} \right>$ is in mass dimension and arises to ensure whether the noncompact model admits de Sitter vacuum or the local Weyl plus $SU(N)$ symmetry is radiatively broken through the one-loop Effective Coleman-Weinberg potential in flat vacua. Followingly, let us assume that the fundamental fields fluctuate around the classical vacuum solutions as follows
\begin{equation}
\begin{aligned}
g_{\mu\nu}&=\bar{g}_{\mu\nu}+\tau h_{\mu\nu}, \qquad \varphi^a =\left< \varphi^a_{vac} \right>+\tau \varphi^a_L,  \qquad \psi^a_{vac}=\tau \psi^a_L, \qquad 
B^a_\mu = \tau B^{a L}_\mu , \qquad C_\mu = \tau C^{aL}_\mu,
\label{vacvalflux}
\end{aligned}
\end{equation}
where $L$ stands for the linearized items. Here, a dimensionless variable $\tau$ is introduced in order to follow the orders throughout expansions. Consequently, by using the fluctuations in Eq.(\ref{vacvalflux}) and quadratic expansion of curvature tensors in Appendix C as well as the following quadratic expansions of excitations 
\begin{equation}
(\varphi^a)^2 = \left< \varphi^a_{vac} \right>^2 \Big [1+2 \tau\frac{\varphi^a_{L}}{\left< \varphi^a_{vac} \right>}+\tau^2\frac{(\varphi^a_L)^2}{\left< \varphi^a_{vac} \right>^2}
+{\cal O}(\tau^3) \Big ], 
\end{equation}
\begin{equation}
(\nabla_\mu A_\nu )^L=\tau \bar{\nabla}_\mu A^{L}_\nu-\tau^2 \Big(\Gamma^{\gamma}_{\mu \nu} \Big)_{L} A^{L}_\gamma+{\cal O}(\tau^3),
\end{equation}
after a straightforward but somewhat long calculation, one finally gets the second order expansion of noncompact Weyl-Einstein-Yang-Mills model in Eq.(\ref{maineqn}) as follows
\begin{equation}
S_{nWEYM}=\bar{S}_{nWEYM}+\tau S^{(1)}_{nWEYM}+\tau^2 S^{(2)}_{nWEYM}+{\cal O}(\tau^3).
\end{equation}
Here ${\cal O}(\tau^0)$ is an unrelated quantity corresponding to types of vacua. On the other side, $S^{(1)}_{nWEYM}$ part reads 
\begin{equation}
{\cal O}(\tau):\hskip 1 cm \Big[2 \alpha \Lambda+\gamma \left< \varphi^a_{vac} \right>^2 \Big] \Big[ \varphi^a_L+\frac{\left< \varphi^a_{vac} \right>}{8} h \Big],
\label{firstord}
\end{equation}
whose solution is dubbed as \emph{vacuum field equation} and characterizes vacuum. Notice that one actually has two alternative candidates for vacuum field equation. To be more precise, setting the first bracket in Eq.(\ref{firstord}) to zero yields 
\begin{equation}
\Lambda=-\frac{\gamma \left< \varphi^a_{vac} \right>^2}{2 \alpha}, 
\label{vacfeper}
\end{equation}
as a vacuum field equation. Observe that this is identical to the one found in Eq.(\ref{vfeqdd}) and relates the classical cosmological constant to VEV of scalar field. Thus, as was shown in the previous section, here only de Sitter vacuum is allowed. On the other side, setting the second bracket in Eq.(\ref{firstord}) to zero gives a completely distinct vacuum field equation: 
\begin{equation}
	\varphi^a_L=-\frac{\left< \varphi^a_{vac} \right>}{8} h,
\end{equation}
which relates the scalar particle to trace of graviton such that they behave in \emph{opposite} directions. In this paper, we take the first choice as vacuum field equation. But, as a future work, it will also be interesting to do the similar stability analysis for the second choice. Finally, the $S^{(2)}_{WnA}$ part which will give the basic quantum oscillators for fluctuations becomes  
\begin{equation}
\begin{aligned}
{\cal O}(\tau^2): \quad &-\frac{\alpha\left< \varphi^a_{vac} \right>^2}{2}h^{\mu\nu} {\cal G}^L_{\mu\nu}+2 \alpha \left< \varphi^a_{vac} \right> \varphi^a_L R^L \\
&+\beta (\partial_\mu \varphi^a_L)^2-8 \alpha \Lambda (\varphi^a_L)^2+\eta \bar{\psi}^L_a \mbox{i} \slashed{\partial} \psi^a_L \\
&+2\sigma (\hat{{\cal C}}^{aL}_{\mu\nu})^2-({\cal M}^{mr}_{C_\mu})^2 C^{mL}_\mu C^{r\mu}_L\\
& +2\sigma (\hat{{\cal B}}^{aL}_{\mu\nu})^2-({\cal M}^{cl}_{B_\mu})^2 B^{cL}_\mu  B^{l\mu} \\
&- 6 \alpha g \left< \varphi^a_{vac} \right>^2 f^{klm} (\bar{\nabla} \cdot C^k_L)\left< \varphi^l_{vac} \right> \left< \varphi^m_{vac} \right>^{-1}\\
&-2 \beta g f^{abc} \varphi^a_L (\bar{\nabla} \cdot B^b_L)\left< \varphi^c_{vac} \right> ,
\label{secordfluc1}
\end{aligned}
\end{equation}
where we made use of Eq.(\ref{vacfeper}) in the derivation. Here, the gauge mass parameters respectively are  
\begin{equation}
({\cal M}^{mr}_{C_\mu})^2 =-g^2 \bigg [\beta f^{alm} f^{apr}-6 \alpha \left< \varphi^a_{vac} \right>^2 f^{klm} f^{npr} \left< \varphi^k_{vac} \right>^{-1}\left< \varphi^n_{vac} \right>^{-1} \bigg ] \left< \varphi^l_{vac} \right>\left< \varphi^p_{vac} \right>, 
\label{gagfldsmasss1}
\end{equation}
\begin{equation}
({\cal M}^{cl}_{B_\mu})^2=-\beta g^2 f^{abc} f^{akl}  \left< \varphi^b_{vac} \right> \left< \varphi^k_{vac} \right>.
\label{gagfldsmasss2}
\end{equation}
Recall that the $3+1$-dimensional linearized curvature tensors in de Sitter backgrounds are defined as follows \cite{DeserTekinPRD, DeserTekinPRL}
\begin{equation}
\begin{aligned}
{\cal G}^L_{\mu\nu}&=R^L_{\mu\nu}-\frac{1}{2} \bar{g}_{\mu\nu} R^L-\Lambda h_{\mu\nu}, \\
R^L&= \bar{\nabla}_\mu  \bar{\nabla}_\nu h^{\mu\nu}-\bar{\square}h-\Lambda h, \\
R^L_{\mu\nu}&=\frac{1}{2}(\bar{\nabla}^\sigma \bar{\nabla}_\mu h_{\sigma\nu}+\bar{\nabla}^\sigma \bar{\nabla}_\nu h_{\sigma\mu}-\bar{\square}h_{\mu\nu}-\bar{\nabla}_\mu \bar{\nabla}_\nu h).
\end{aligned}
\end{equation}
To be able to obtain the particle spectrum about the vacua, one naturally needs the isolated harmonic oscillator of each excitation. However, as is manifest in Eq.(\ref{secordfluc1}), the oscillators are coupled among themselves and therefore they are not at the desired forms yet. So, one has somehow to decouple the excitations from each others: as was done in \cite{Tanhayindim, TanhayiUnNMG}, the term involving coupling of scalar and gauge fields can actually be decoupled via a reasonable choice of gauge-fixing condition which will also eliminate the non-dynamical DOF. On the other side, the term that contains coupling between scalar and tensor fluctuations can be decoupled via a suitable redefinition of tensor perturbation. Thus, let us proceed accordingly: 

\subsection*{Decoupling Excitations}
\subsubsection{Gauge-Choice}  
As is well-known, gauging system generally comes with non-propagating DOF which have to be ruled out via fixing gauge. In our case, in addition to elimination of existing redundancies, it is also expected from gauge choice to decouple gauge and scalar excitations from each others at the linearized level. For this purpose, let us note that in accordance with the Sec.II and Sec.III, one could assume the noncompact gauge covariant derivative of noncompact gauge field to be  
\begin{equation}
{\cal D}_\mu A_\nu=\partial_\mu A_\nu-\mbox{i}g[A_\mu, A_\nu],
\label{noncompgagcovderggafix}
\end{equation}
with which one can further allow the incorporation of spacetime and noncompact gauge covariant derivative to be 
\begin{equation}
{\cal D}_\mu A_\nu=\nabla_\mu A_\nu-\mbox{i}g[A_\mu, A_\nu].
\label{noncompgagcovdergafix}
\end{equation}
Thereafter, it is straightforward to show that Eq.(\ref{noncompgagcovdergafix}) transforms according to adjoint representation of $SL(N, \mathbb{C})$ as follows
\begin{equation}
({\cal D}_\mu A^\mu )^{'}={\cal U} ({\cal D}_\mu A^\mu) {\cal U}^{-1}+\nabla^\mu \Big [{\cal U} A_\mu {\cal U}^{-1}+\frac{1}{\mbox{i}g} (\partial_\mu {\cal U}){\cal U}^{-1} \Big]-{\cal U} \nabla_\mu A^\mu {\cal U}^{-1},
\end{equation}
which, with the following fixing (see below for a brief discussion on this setting)
\begin{equation}
\nabla^\mu \Big [{\cal U} A_\mu {\cal U}^{-1}+\frac{1}{\mbox{i} g} (\partial_\mu {\cal U}){\cal U}^{-1} \Big]-{\cal U} \nabla_\mu A^\mu {\cal U}^{-1}=0,
\label{gaugfixzeroset}
\end{equation}
reduces to
\begin{equation}
({\cal D}_\mu A^\mu )^{'}={\cal U} ({\cal D}_\mu A^\mu) {\cal U}^{-1}.
\end{equation}
Thus, one can in fact take the Lorentz-like gauge fixing term
\begin{equation}
{\cal D}_\mu A^\mu=\nabla_\mu A^\mu=0,
\label{noncompggagcovdergafix}
\end{equation}
as a noncompact gauge-fixing condition which, at the linearized level, reduces to the background covariant Lorentz-like condition
\begin{equation}
\bar{\nabla} \cdot A^L=0.
\label{gagfixnoncomp}
\end{equation}
Observe that, due to $A^L_\mu=B^L_\mu+\mbox{i} C^L_\mu$, Eq.(\ref{gagfixnoncomp}) simultaneously eliminates unphysical DOF associated to both Weyl's scale and $SU(N)$ phase symmetries respectively via
\begin{equation}
\bar{\nabla} \cdot B^L=0, \qquad \bar{\nabla} \cdot C^L=0.
\label{gaugfixnncmpct}
\end{equation}
Apparently, with these gauge-fixing conditions, the terms that contain coupling between scalar and vector particles in Eq.(\ref{secordfluc1}) automatically drop out. Needless to also say that Eq.(\ref{gaugfixzeroset}) is actually the noncompact generalization of ordinary harmonic gauge, i.e., $\bar{\square} \zeta(x)=0$. In order to see this, let us notice that by expanding the noncompact group transformation operator in generator bases of daughter compact $SU(N)$ as 
\begin{equation}
{\cal U}=e^{-\mbox{i} [w^a(x)+\mbox{i} \theta^a(x) ] T^a},
\end{equation}
Eq.(\ref{gaugfixzeroset}) turns into 
\begin{equation}
\Big( \square w^a+g f^{abc} A^b_\mu \partial^\mu w^c \Big)+\mbox{i} \Big(\square \theta^a+g f^{abc} A^b_\mu \partial^\mu \theta^c \Big)=0,
\end{equation}
which yields the harmonic gauge-type extensions for Weyl's scale and $SU(N)$ phase parts respectively as follows
\begin{equation}
	\square w^a+g f^{abc} A^b_\mu \partial^\mu w^c=0, \hskip 1cm \square \theta^a+g f^{abc} A^b_\mu \partial^\mu \theta^c=0, 
\end{equation}
that, at the linearized level, become
\begin{equation}
\bar{\square} w^a=0, \hskip 1 cm \bar{\square} \theta^a=0.	
\end{equation}

\subsubsection{Redefining Graviton Field}
On the other side, one needs to redefine the tensor fluctuation in order to decouple it from scalar excitation. More precisely, one can easily show that with the following redefinition of graviton field
\begin{equation}
h_{\mu\nu}=\hat{h}_{\mu\nu}-2  \left< \varphi^a_{vac} \right>^{-1} \, \bar{g}_{\mu\nu} \varphi^a_L,
\label{redmetric}
\end{equation}
the fundamental linearized curvature tensors change to 
\begin{equation}
\begin{aligned}
R^L&=\hat{R}^L+6 \left< \varphi^a_{vac} \right>^{-1} \Big(\bar{\square} \varphi^a_L+\frac{4}{3} \varphi^a_L \Big), \\
R^L_{\mu\nu}&=\hat{R}^L_{\mu\nu}+2 \left< \varphi^a_{vac} \right>^{-1} \Big( \bar{\nabla}_\mu  \bar{\nabla}_\nu \varphi^a_L+\frac{1}{2} \bar{g}_{\mu\nu} \bar{\square} \varphi^a_L\Big),\\
{\cal G}^L_{\mu\nu}&= \hat{{\cal G}}^L_{\mu\nu}+2 \left< \varphi^a_{vac} \right>^{-1} \Big( \bar{\nabla}_\mu \bar{\nabla}_\nu \varphi^a_L-\bar{g}_{\mu\nu}\bar{\square} \varphi^a_L-4\Lambda \bar{g}_{\mu\nu} \varphi^a_L \Big),\\
h^{\mu\nu} {\cal G}^L_{\mu\nu}&= \hat{h}^{\mu\nu} \hat{{\cal G}}^L_{\mu\nu}+12 \left< \varphi^a_{vac} \right>^{-2}  \Big( \frac{\left< \varphi^a_{vac} \right>}{3} \varphi^a_L \hat{R}^L+ \varphi^a_L \bar{\square} \varphi^a_L+\frac{4 \Lambda}{3} (\varphi^a_L)^2 \Big),
\end{aligned}
\end{equation}
with which the scalar and graviton fluctuations eventually decouple from each others.

Thus, as was emphasized above, substituting the redefinition of tensor fluctuation and background gauge-fixing condition at the linearized level into Eq.(\ref{secordfluc1}) decouples all the fundamental quantum oscillators, and one is finally left with a completely isolated system
\begin{equation}
\begin{aligned}
{\cal O}(\tau^2): \quad &-\frac{\alpha\left< \varphi^a_{vac} \right>^2}{2} \hat{h}^{\mu\nu} \hat{{\cal G}}^L_{\mu\nu} -\frac{1}{2}(12 \alpha-2\beta) (\partial_\mu \varphi^a_L)^2+\eta \bar{\psi}^L_a \mbox{i} \slashed{\partial} \psi^a_L  \\
&+2\sigma (\hat{{\cal C}}^{aL}_{\mu\nu})^2-({\cal M}^{mr}_{C_\mu})^2 C^{mL}_\mu C^{r\mu}_L +2\sigma (\hat{{\cal B}}^{aL}_{\mu\nu})^2-({\cal M}^{cl}_{B_\mu})^2 B^{cL}_\mu  B^{l\mu},
\label{secordfluc2}
\end{aligned}
\end{equation}
where the gauge mass parameters are given in Eq.(\ref{gagfldsmasss1}) and Eq.(\ref{gagfldsmasss2}). Here, the first term is the linearized Einstein's gravity with an effective Newton's constant generated from VEV of scalar fields. Therefore, the model propagates with a \emph{unitary} massless graviton as long as $\alpha > 0$. The second term describes a kinetic term for each scalar particle corresponding to each generator basis of $SU(N)$. Hence, due to the number of generators, the model actually propagates with $N^2-1$ \emph{non-ghosts} massless scalar bosons for $\beta < 6 \alpha $. Notice that with this parameter region, the unitarities of scalar bosons and gravitons are linked to each others. Followingly, the last term in the first line is the Dirac theory. Thus, due to the type of representation, the model has $N$ massless Dirac fermions.  Finally, the terms in the second line are Proca-type Lagrangians. Accordingly, by setting the corresponding kinetic terms to their canonically normalized values, (i.e., $\sigma =-1/8 $), the model will propagate with \emph{non-tachyonic} $N^2-1$ massive Abelian and non-Abelian gauge bosons with masses ${\cal M}^{mr}_{C_\mu}$ and ${\cal M}^{cl}_{B_\mu}$ for the following parameters region
\begin{equation}  
\beta <0< 6 \alpha. 
\end{equation}  
Observe that the unitarities of gauge fields do \emph{not} allow to take the critical choice $\beta = 6 \alpha $ at which the scalar particles would completely disappear among the model. Note also that the mass of gauge field corresponding to a specific generator basis interestingly gets contributions from the ones corresponding to other generator bases.

\section{conclusion}         
By supposing the Higgs-like field to be the source of transition from Einstein's geometry to Weyl's geometry, we have formulated a $3+1$-dimensional \emph{noncompact} Weyl-Einstein-Yang-Mills model. The model physically defines a local $SL(N, \mathbb{C})$ [i.e., Weyl's scale plus $SU(N)$ phase] invariant Higgs-like field, conformally coupled to a generic Weyl-invariant dynamical background. By construction, the Higgs-like sector produces the Weyl's conformal invariance of system. To completely generate the Weyl's symmetry from Higgs-like sector, we have initially chosen the Higgs-like field to be an element of $SU(N)$ in the adjoint representation in analogy with the $SU(5)$ Georgi-Glashow model. Thereupon, we have smoothly extended the local gauge group of adjoint Higgs-like field [that is, the compact $SU(N)$] via a viable embedding of Weyl's symmetry. This integration of symmetries has resulted in the noncompact $SL(N, \mathbb{C})$. Finally, we have constructed the Weyl's scalar and gauge fields from proper superpositions of fundamental elements of $SL(N, \mathbb{C})$ expanded in the generator bases of $SU(N)$. The model does not involve any dimensionful parameters and genuine presence of de Sitter vacuum spontaneously breaks the noncompact gauge symmetry in an analogous manner to the Higgs mechanism. As to flat vacuum, the dimensionful parameter emerges within the dimensional transmutation in quantum field theories, and so the symmetry is radiatively broken through the one-loop Coleman-Weinberg Effective potential. After the symmetry is broken, all the interaction parameters freeze to constant values. Here, we have demonstrated that the mere expectation of arriving at Einstein's gravity in the broken phases prevents anti-de Sitter space to be its stable constant curvature vacuum. Additionally, we have also performed the tree-level perturbative unitarity of model in de Sitter space. We have shown that the model is unitary in de Sitter and flat vacua around which a massless graviton, $N^2-1$ massless scalar bosons, $N$ massless Dirac fermions, $N^2-1$ Proca-type massive Abelian and non-Abelian gauge bosons are generically propagated. Here, all the corresponding lower spin fields in all the generator bases give contribution to the Weyl's conformal scaling of system. In the perturbative study, we have also noted that there are two distinct candidates for vacuum field equation: in the first option, the cosmological constant and VEV of scalar fields are linked to each others, whereas, in the second option, the scalar bosons are linked to trace of graviton in such a way that the scalar bosons all together yield a repulsive force in an analogous manner with the Dark energy.

As for the coupling of matter to the model, due to the main purpose of constructing a well-behaved larger geometrical description of gravitational force at first, here we have not taken this sector into account. However, let us briefly note the following points: recall that if one wants a full model (i.e., geometry plus matter sectors) to be Weyl-invariant, the matter sector should also be Weyl-invariant. In this case, the vacuum solution may spontaneously break the symmetry, whereas if the symmetry is explicitly broken, the theory is then no longer Weyl-invariant. One can also couple a Weyl-non-invariant matter to a Weyl-invariant gravity. This will simply mean that the whole model is not Weyl-invariant even if its geometry sector is. Of course, these are the common information about the Weyl invariance and require a detailed analysis in order to explicitly see what contribution(s) these particular cases bring. For future, in addition to the standard cosmological analyses of the $3+1$-dimensional \emph{noncompact} Weyl-Einstein-Yang-Mills model, it also seems to be appealing to study it in the context of holography to find out its complete asymptotic symmetry algebra, as in the \cite{Grumiller} in which it is shown that, under certain boundary conditions on the metric, the asymptotic symmetry algebra of $2+1$-dimensional conformal Chern-Simons gravity without matter field turns out to be the conformal algebra plus a $u(1)_k$ current.

\section{\label{ackno} Acknowledgments}

We would like to thank Roman Jackiw, Bayram Tekin, Daniel Grumiller and Tuna Yildirim for several useful discussions and suggestions. We would also like to thank a very conscientious referee for his/her constructive remarks and advices on the paper. We finally want to thank Ibrahim Burak Ilhan and Ercan Kilicarslan for critical readings of the paper and Marta Campigotto for informing their interesting paper. This work is supported by TUBITAK 2219 Postdoctoral Scholarship.

\section{Appendix A: Field Equations}           

As was stated in Sec.III, one naturally needs to find the field equations of $3+1$ dimensional noncompact Weyl-Einstein-Yang-Mills model in order to see if the vacuum breaks the noncompact $SL(N, \mathbb{C})$ [that is, Weyl plus $SU(N)$] gauge symmetry. For this purpose, we give the corresponding field equations in this section. Therefore, let us first note that by varying Eq.(\ref{maineqn}) with respect to $g^{\mu\nu}$, up to a boundary term, one arrives at
 \begin{equation}
 \begin{aligned}
 &\alpha (\varphi^a)^2 \bigg[ G_{\mu\nu}+3g g_{\mu\nu} \Big[ f^{klm} (\nabla_\alpha C^{k\alpha}) \varphi^l (\varphi^m)^{-1} +g f^{klm} f^{npr} C^k_\alpha \varphi^l (\varphi^m)^{-1}\,  C^{n\alpha} \varphi^p (\varphi^r)^{-1} \Big]\\
 &\hskip 1cm -6g \Big[ f^{klm} (\nabla_\mu C^k_\nu) \varphi^l (\varphi^m)^{-1} +g f^{klm} f^{npr}  C^k_\mu \varphi^l (\varphi^m)^{-1}\, C^n_\nu \varphi^p (\varphi^r)^{-1} \Big] \bigg] \\
&+ \eta (\varphi^a)^2 \Big[ \bar{\psi}_b \mbox{i} \gamma_\nu ({\cal D}_\mu \psi)^b -\frac{1}{2} g_{\mu\nu}\bar{\psi}_b \mbox{i} (\slashed{\cal D}\psi)^b \Big] \\
 &+\beta \bigg[({\cal D}_\mu \varphi^a)({\cal D}_\nu \varphi^a)^{+}-\frac{1}{2}g_{\mu\nu} \Big[ ({\cal D}_\rho \varphi^a)({\cal D}^\rho \varphi^a)^{+}+\frac{\gamma}{\beta}(\varphi^a)^4 \Big] \bigg]\\
 &-4 \sigma \bigg \{ {\cal B}^a_{\mu\alpha}{\cal B}^a_{\nu}{^\alpha}+{\cal C}^a_{\mu\alpha}{\cal C}^a_{\nu}{^\alpha}-\frac{1}{4} g_{\mu\nu} \Big ({\cal B}^a_{\alpha \beta}{\cal B}^{a\alpha\beta}+{\cal C}^a_{\alpha\beta}{\cal C}^{a\alpha\beta} \Big ) \\
&\hskip 0.8 cm -\frac{2\Theta^a \Theta^b}{{\bf \Theta}^2}\Big[ {\cal B}^a_{\mu\alpha}{\cal B}^b_{\nu}{^\alpha}+{\cal C}^a_{\mu\alpha}{\cal C}^b_{\nu}{^\alpha}-\frac{1}{4} g_{\mu\nu} \Big ({\cal B}^a_{\alpha \beta}{\cal B}^{b\alpha\beta}+{\cal C}^a_{\alpha\beta}{\cal C}^{b\alpha\beta} \Big ) \Big]  \bigg\}=0,
 \label{feqofmetrcs}
 \end{aligned}	
 \end{equation}
 where $G_{\mu\nu}=R_{\mu\nu}-\frac{1}{2} g_{\mu\nu}R$ is the celebrated Einstein tensor. Secondly, by varying Eq.(\ref{maineqn}) with respect to $ \varphi^a $, one gets   
 \begin{equation}
 \begin{aligned}
 & 2 \alpha \varphi^a \Big [ R- 6g f^{klm} (\nabla_\mu C^{k\mu}) \varphi^l (\varphi^m)^{-1}-6 g^2 f^{klm} f^{npr} C^k_\mu \varphi^l (\varphi^m)^{-1}\, C^{n\mu} \varphi^p (\varphi^r)^{-1} \Big]\\
 &+6\alpha g (\varphi^m)^2 \Big [f^{abc} \nabla_\mu C^{b\mu} (\varphi^c)^{-1}+ (\varphi^a)^{-2} f^{abc} (\nabla_\mu C^{b\mu}) \varphi^c+ 2 g f^{abc}f^{klm} C^b_\mu  (\varphi^{c})^{-1}  C^{k\mu} \varphi^l (\varphi^m)^{-1} \\
 & \hskip 2cm + 2 g(\varphi^{a})^{-2} f^{abc}f^{klm} C^b_\mu \varphi^c C^{k\mu} \varphi^l (\varphi^m)^{-1} \Big]\\
 & -2 \beta \Big[\square \varphi^a - g f^{abc} \partial_\mu B^{b\mu} \varphi^c + g^2 f^{abc} f^{ckl} ( B^b_\mu B^{k\mu}+C^b_\mu C^{k\mu} ) \varphi^l-\frac{2 \gamma}{\beta} (\varphi^b)^2 \varphi^a \Big]\\
&+2 \eta \Big[\varphi^a \bar{\psi}_b \mbox{i} (\slashed{\cal D} \psi)^b+\frac{1}{2} (\varphi^c)^2 \bar{\psi}\mbox{i} \gamma^\mu \frac{\delta \Theta}{\delta \varphi^a} {\cal D}_\mu \psi \Big]+4 \sigma \Big [\frac{2 \frac{\delta \Theta^c }{\delta \varphi^a}\Theta^b}{{\bf \Theta}^2}- \frac{ \frac{\delta \Theta^2 }{\delta \varphi^a}\Theta^c \Theta^b}{{\bf \Theta}^3}\Big]F^{+c}_{\mu\nu}F^{b\mu\nu}=0.
 \label{feqscfld}
 \end{aligned}
 \end{equation}
The fermionic field equation reads
\begin{equation}
\mbox{i} (\varphi^b)^2 \Gamma^\mu ({\cal D}_\mu \psi)^a=0
\end{equation}
Thirdly, variation with respect to the non-Abelian gauge field $B^{a\mu}$ gives 
 \begin{equation}
 \begin{aligned}
&\eta (\varphi^b)^2 \bar{\psi}_a \Gamma_\mu \psi^a+ g \beta f^{abc} \Big[\partial_\mu \varphi^b+g  f^{bkl} B^k_\mu \varphi^l \Big] \varphi^c\\
&-8 \sigma \bigg\{\partial^\nu {\cal B}^a_{\mu\nu}-g f^{abc} \Big({\cal B}^b_{\mu\nu} B^{c\nu}+{\cal C}^b_{\mu\nu} C^{c\nu} \Big) -2 \partial_\nu \Big (\frac{\Theta^a \Theta^b}{{\bf \Theta}^2} {\cal B}^{b\mu\nu} \Big)+2g \frac{\Theta^m \Theta^b}{{\bf \Theta}^2} f^{amn} \Big( {\cal B}^b_{\mu\nu} B^{n\nu}+{\cal C}^b_{\mu\nu} C^{n\nu}\Big)\bigg\}=0.
 \label{feqnonabefld}
 \end{aligned}
 \end{equation}
 Finally, the gauge covariant field $C^{a\mu}$ part yields 
 \begin{equation}
 \begin{aligned}
 & 12 \alpha g (\varphi^d)^2 \Big [ (\varphi^d)^{-1} \partial_\mu \varphi^d \, f^{abc} \varphi^b (\varphi^c)^{-1}+\frac{1}{2}f^{abc} (\partial_\mu \varphi^b) (\varphi^c)^{-1}+ \frac{1}{2} f^{abc} \varphi^b \partial_\mu (\varphi^c)^{-1}  \\
 & \hskip 2cm-g f^{abc}f^{klm} \varphi^b (\varphi^c)^{-1} C^{k\mu} \varphi^l (\varphi^m)^{-1} \Big] \\
 &-\beta g \Big[\partial_\mu (\varphi^d)^2 f^{abc} \varphi^b (\varphi^c)^{-1}  + 2 g f^{abc} f^{blm} \varphi^c C^l_\mu \varphi^m \Big] +g\eta (\varphi^b)^2 \bar{\psi}\mbox{i} \Gamma_\mu \psi\\
&-8\sigma \bigg \{\partial^\nu {\cal C}^a_{\mu \nu}-g f^{abc} ({\cal C}^b_{\mu \nu} B^{c\nu} -{\cal B}^b_{\mu \nu} C^c_\nu ) -2 \Big[  \partial^\nu \Big(\frac{\Theta^a \Theta^b}{{\bf \Theta}^2} {\cal C}^b_{\mu\nu} \Big)+g f^{abn} \frac{\Theta^m \Theta^b}{{\bf \Theta}^2} ({\cal B}^m_{\mu\nu}C^{n\nu}- {\cal C}^m_{\mu\nu}B^{n\nu} ) \Big] \bigg\}=0.
 \label{feqabefld}
 \end{aligned}
 \end{equation}
Notice that the complicated field equations are expected because the model comprises more than one fields, non-minimally coupled to each others. Here, with the identities among structure constants in the literature, the field equations can be further recast to include such as the Casimir operators of adjoint representation. But, since we want to illustrate how the fields come together explicitly and our primary aim is to find the corresponding symmetry breaking mechanisms for de Sitter and flat vacua, we will not do this and leave them in the above forms.  
 
\section{Appendix B: Adjoint Higgs-like field of $SU(N)$}                
Since the Higgs-like field of $SU(N)$ in the adjoint representation plays pivotal role in the construction of $3+1$-dimensional noncompact Weyl-Einstein-Yang-Mills model, we briefly review its properties in this section. To do so, let us first recall that as one goes beyond the Standard Model, one naturally needs to define a suitable larger gauge group in order to shed light on these extreme energy regimes. In this respect, countless number of models have been introduced so far. Among these theories, Grand Unified Theories of fundamental particles are particularly interesting because they provide physically more prescient models: as is known, these theories try to reconcile the coupling constants of Strong, Weak and Electromagnetic interactions via larger gauge groups which are expected to be broken towards the Standard Model throughout successive symmetry breaking. In this family, the Georgi-Glashow model which comprises $SU(5)$ gauge group is specifically much more interesting because it supplies the simplest unified gauge theory and does not require any extra matter fields other than the ones in Standard Model \cite{Glashw}\footnote{See for example \cite{Romao} for a comprehensive review of the Georgi-Glashow model.}. Remember that here the Higgs-like sector are defined in two representations (that is, the adjoint and fundamental representations that admit $24$ and $5$ elements, respectively) which correspondingly trigger the following spontaneous chain symmetry breaking
\begin{equation}
SU(5) \rightarrow SU(3)_C \times SU(2)_I \times U(1)_Q \rightarrow SU(3)_C \times U(1)_Q, 
\label{ssbkggm}   
\end{equation}
via the Coleman-Weinberg mechanism \cite{GGSBCM1, GGSBCM2, GGSBCM3, GGSBCM4, GGSBCM5, GGSBCM6}. Here, $SU(3)_C$, $SU(2)_I$ and $U(1)_Q$ are the color, isospin and hyper-charge gauge groups, respectively. It is this first kind of Higgs-like field that we benefit in the formulation of noncompact model. Hence, let us now shortly examine some of its basics for the generic case: recall that the Higgs-like field of $SU(N)$ in the adjoint representation transforms according to
\begin{equation}  
 \varphi \rightarrow \varphi^{'}=U \varphi U^{-1},
 \label{trnsofcompsclf}
 \end{equation}
 where $U$ is the corresponding group transformation. Here, one is allowed to expand $\varphi$ in the generator bases as follows
 \begin{equation}
 \varphi= \varphi^a T^a, \qquad a=1, \cdots N^2-1,
 \label{realscalcomp}
 \end{equation}
 in which $T^a$ denotes the generators of symmetry group and $\varphi^a$ is the real scalar field corresponding to the magnitude of Higgs-like field in $a^{th}$ generator basis. As to the local $SU(N)$ invariant scalar field theory, one needs to replace the ordinary derivative with the gauge covariant derivative in the adjoint representation: 
 \begin{equation}
 \hat{D}_\mu \varphi = \partial_\mu \varphi-\mbox{i} g [B_\mu, \varphi].
 \label{gagcovcompgrs}
 \end{equation}
 Notice that by using $[T^a, T^b]=\mbox{i} f^{abc} T^{c}$ where $ f^{abc} $ refers to structure constant, Eq.(\ref{gagcovcompgrs}) can also be projected in the generator bases as $ \hat{D}_\mu \varphi =\hat{D}_\mu \varphi^a T^a $ in which the real part reads
 \begin{equation}
 \hat{D}_\mu \varphi^a=\partial_\mu \varphi^a+ g f^{abc} B^b_\mu \varphi^c.
 \label{gagcompreal}
 \end{equation}
 Here, the compensating non-Abelian gauge field $ B_\mu $ changes according to the adjoint representation as follows
 \begin{equation}
 B_\mu  \rightarrow  B^{'}_\mu=U B_\mu U^{-1}+\frac{1}{\mbox{i} g} (\partial_\mu U) U^{-1},
 \label{trfgagfildcompf}
 \end{equation}
 with whom Eq.(\ref{gagcovcompgrs}) transforms as
 \begin{equation}
 \hat{D}_\mu \varphi \rightarrow \hat{D}_\mu \varphi^{'}=U (\hat{D}_\mu \varphi) U^{-1}.
 \label{trasfrofgcdcopms}
 \end{equation}
 Meanwhile, one naturally needs to consider the following field-strength tensor
 \begin{equation}
 \hat{F}_{\mu\nu}=\partial_\mu B_\nu-\partial_\nu B_\mu-\mbox{i} g [B_\mu,B_\nu] \hskip 1 cm \mbox{with} \hskip 1 cm \hat{F}_{\mu\nu} \rightarrow \hat{F}_{\mu\nu}=U\hat{F}_{\mu\nu} U^{-1},
 \label{nonablnfts}
 \end{equation}
 and construct the corresponding Yang-Mills kinetic term in order to provide dynamics to the gauge fields. Note that Eq.(\ref{nonablnfts}) can also be recast as $ \hat{F}_{\mu\nu} =\hat{F}^a_{\mu\nu} T^a$ where 
 \begin{equation}
 \hat{F}^a_{\mu\nu}=\partial_\mu B^a_\nu-\partial_\nu B^a_\mu+g f^{abc} B^b_\mu B^c_\nu.
 \end{equation}

\section{Appendix C: Second Order Expansion of Curvature Tensors}    
In this section, we give the quadratic expansion of fundamental curvature tensors about an $n$-dimensional arbitrary curved background \cite{TanhayiUnNMG, Gullu}. To do so, let us first notice that an arbitrary spacetime with $g_{\mu\nu}$ can be split into a background spacetime $\bar{g}_{\mu\nu}$ plus a sufficiently small fluctuation $h_{\mu\nu}$ as follows  
\begin{equation}
g_{\mu \nu}=\bar{g}_{\mu \nu}+\tau h_{\mu \nu},
\label{metrflct}
\end{equation}
where $\tau$ is an expansion tracking dimensionless variable. Here, all the tensor operations are taken place via the background metric. In accordance with Eq.(\ref{metrflct}), the second order expansions of dual metric and volume element respectively become
\begin{equation}
g^{\mu \nu}=\bar{g}^{\mu \nu}-\tau h^{\mu \nu}+\tau^2 h^{\mu \rho} h^\nu_\rho+ {\cal O}(\tau^3), \qquad \sqrt{-g} =\sqrt{-\bar{g}} \bigg[1+\frac{\tau}{2}h+\frac{\tau^2}{8} \Big (h^2-2 h^2_{\mu \nu} \Big)+{\cal O}(\tau^3) \bigg].
\label{invmetrflct}
\end{equation}
Then, by using Eq.(\ref{metrflct}) and Eq.(\ref{invmetrflct}), one obtains the quadratic expansion of Levi-Civita connection as follows
\begin{equation}
\Gamma^\rho_{\mu \nu}=\bar{\Gamma}{^\rho_{\mu \nu}}+ \tau \Big (\Gamma^\rho_{\mu \nu} \Big)_{L}
-\tau^2 h^\rho_\beta \Big (\Gamma^\beta_{\mu \nu} \Big)_{L} +{\cal}(\tau^3),
\label{qdrctlct}
\end{equation}
where $ \bar{\Gamma}{^\rho_{\mu \nu}} $ is the relevant background connection. Here, the linearized part reads
\begin{equation}
\Big (\Gamma^\rho_{\mu \nu} \Big)_{L}=\frac{1}{2}\bar{g}^{\rho \lambda} \Big (\bar{\nabla}_\mu h_{\nu \lambda}+ \bar{\nabla}_\nu
h_{\mu \lambda}-\bar{\nabla}_\lambda h_{\mu \nu} \Big ),
\end{equation}
with which the second order expansion of Riemann tensor becomes
\begin{equation}
\begin{aligned}
R{^\mu}{_{\nu \rho \sigma}}=& \bar{R}{^\mu}{_{\nu \rho \sigma}}+ \tau \Big (R{^\mu}{_{\nu \rho \sigma}} \Big)_{L}
-\tau^2 h^\mu_\beta \Big ( R{^\beta}{_{\nu \rho \sigma}} \Big )_{L} \\
&\hskip 1 cm-\tau^2 \bar{g}^{\mu \alpha} \bar{g}_{\beta \gamma} \bigg [\Big
(\Gamma^\gamma_{\rho \alpha} \Big)_{L} \Big(\Gamma^\beta_{\sigma
	\nu} \Big)_{L}- \Big (\Gamma^\gamma_{\sigma \alpha} \Big)_{L}
\Big(\Gamma^\beta_{\rho \nu} \Big)_{L} \bigg ]+{\cal O}(\tau^3),
\label{qdrcremtns}
\end{aligned}
\end{equation}
where 
\begin{equation}
\Big ( R{^\beta}{_{\nu \rho \sigma}} \Big )_{L}= \frac{1}{2} \Big
(\bar{\nabla}_\rho \bar{\nabla}_\sigma h^\mu_\nu+\bar{\nabla}_\rho
\bar{\nabla}_\nu h^\mu_\sigma-\bar{\nabla}_\rho \bar{\nabla}^\mu
h_{\sigma \nu}-\bar{\nabla}_\sigma \bar{\nabla}_\rho
h^\mu_\nu-\bar{\nabla}_\sigma \bar{\nabla}_\nu
h^\mu_\rho+\bar{\nabla}_\sigma \bar{\nabla}^\mu h_{\rho \nu} \Big
). \label{riem}
\end{equation}
 Subsequently, the quadratic expansion of Ricci tensor reads
\begin{equation}
\begin{aligned}
R_{\nu \sigma}=& \bar{R}_{\nu \sigma}+\tau \Big (R_{\nu \sigma} \Big)_{L}-\tau^2 h^\mu_\beta \Big(R^\beta{_{\nu \mu \sigma}} \Big)_{L} \\
& \hskip 0.7 cm - \tau^2 \bar{g}^{\mu \alpha} \bar{g}_{\beta \gamma} \bigg [\Big
(\Gamma^\gamma_{\mu \alpha} \Big)_{L} \Big(\Gamma^\beta_{\sigma
	\nu} \Big)_{L}- \Big (\Gamma^\gamma_{\sigma \alpha} \Big)_{L}
\Big(\Gamma^\beta_{\mu \nu} \Big)_{L} \bigg ]+{\cal O}(\tau^3).
\label{linrzriccitsr}
\end{aligned}
\end{equation}
Here
\begin{equation}
R^{L}_{\nu \sigma}=\frac{1}{2} \Big (\bar{\nabla}_\mu \bar{\nabla}_\sigma h^\mu_\nu+\bar{\nabla}_\mu \bar{\nabla}_\nu
h^\mu_\sigma- \bar{\Box}h_{\sigma \nu}-\bar{\nabla}_\sigma \bar{\nabla}_\nu h \Big).
\label{ricc}
\end{equation}
Finally, the second order expansion of Ricci scalar becomes    
\begin{equation}
\begin{aligned}
R=\bar{R}+\tau R_{L}+\tau^2 \bigg \{& \bar{R}^{\rho \lambda}h_{\alpha \rho}h^\alpha_\lambda-h^{\nu \sigma}
\Big(R_{\nu \sigma} \Big)_{L}-\bar{g}^{\nu \sigma} h^\mu_\beta \Big(R^\beta{_{\nu \mu \sigma}} \Big)_{L} \\
& -\bar{g}^{\nu \sigma} \bar{g}^{\mu \alpha} \bar{g}_{\beta
	\gamma} \bigg [\Big (\Gamma^\gamma_{\mu \alpha} \Big)_{L}
\Big(\Gamma^\beta_{\sigma \nu} \Big)_{L}- \Big
(\Gamma^\gamma_{\sigma \alpha} \Big)_{L} \Big(\Gamma^\beta_{\mu
	\nu} \Big)_{L} \bigg ] \bigg \}+{\cal O}(\tau^3),
\end{aligned}
\end{equation}
where 
\begin{equation}\label{rl}
R^{L}=(g^{\alpha \beta} R_{\alpha \beta})^{L}=\bar{g}^{\alpha \beta} R^{L}_{\alpha \beta}-\bar{R}^{\alpha \beta} h_{\alpha \beta}.
\end{equation}
Thus, with all the above settings, the linearized Einstein tensor reads
\begin{equation}
 {\cal G}_{\mu\nu}^L=R_{\mu\nu}^L-\frac{1}{2}\bar{g}_{\mu\nu}R^L-\frac{2\Lambda}{n-2} h_{\mu\nu}. 
\end{equation}

\end{document}